\documentclass[twocolumn,secnumarabic,amssymb, nobibnotes, aps, prl, superscriptaddress, longbibliography]{revtex4-2}

\usepackage{graphicx}
\usepackage{natbib}
\usepackage{xcolor}
\usepackage[colorlinks,allcolors=blue]{hyperref}

\begin{document}

\title{Semi-Dirac transport and anisotropic localization in polariton honeycomb lattices}%

\author{B.~Real}
\affiliation{Univ. Lille, CNRS, UMR 8523 -- PhLAM -- Physique des Lasers Atomes et Mol\'ecules, F-59000 Lille, France}

\author{O.~Jamadi}
\affiliation{Univ. Lille, CNRS, UMR 8523 -- PhLAM -- Physique des Lasers Atomes et Mol\'ecules, F-59000 Lille, France}

\author{M.~Mili\'cevi\'c}
\affiliation{Universit\'e Paris-Saclay, CNRS, Centre de Nanosciences et de Nanotechnologies, 91120, Palaiseau, France}

\author{N.~Pernet}
\affiliation{Universit\'e Paris-Saclay, CNRS, Centre de Nanosciences et de Nanotechnologies, 91120, Palaiseau, France}

\author{P.~St-Jean}
\affiliation{Universit\'e Paris-Saclay, CNRS, Centre de Nanosciences et de Nanotechnologies, 91120, Palaiseau, France}

\author{T.~Ozawa}
\affiliation{Advanced Institute for Materials Research, Tohoku University, Sendai 980-8577, Japan}
\affiliation{Interdisciplinary Theoretical and Mathematical Sciences Program (iTHEMS), RIKEN, Wako, Saitama 351-0198, Japan}

\author{G.~Montambaux}
\affiliation{Universit\'e Paris-Saclay, CNRS, Laboratoire de Physique des Solides, 91405, Orsay, France}

\author{I.~Sagnes}
\affiliation{Universit\'e Paris-Saclay, CNRS, Centre de Nanosciences et de Nanotechnologies, 91120, Palaiseau, France}

\author{A.~Lema\^itre}
\affiliation{Universit\'e Paris-Saclay, CNRS, Centre de Nanosciences et de Nanotechnologies, 91120, Palaiseau, France}

\author{L.~Le Gratiet}
\affiliation{Universit\'e Paris-Saclay, CNRS, Centre de Nanosciences et de Nanotechnologies, 91120, Palaiseau, France}

\author{A.~Harouri}
\affiliation{Universit\'e Paris-Saclay, CNRS, Centre de Nanosciences et de Nanotechnologies, 91120, Palaiseau, France}

\author{S.~Ravets}
\affiliation{Universit\'e Paris-Saclay, CNRS, Centre de Nanosciences et de Nanotechnologies, 91120, Palaiseau, France}

\author{J.~Bloch}
\affiliation{Universit\'e Paris-Saclay, CNRS, Centre de Nanosciences et de Nanotechnologies, 91120, Palaiseau, France}

\author{A.~Amo}
\affiliation{Univ. Lille, CNRS, UMR 8523 -- PhLAM -- Physique des Lasers Atomes et Mol\'ecules, F-59000 Lille, France}

\begin{abstract} 

Compression dramatically changes the transport and localization properties of graphene. This is intimately related to the change of symmetry of the Dirac cone when the particle hopping is different along different directions of the lattice. In particular, for a critical compression, a semi-Dirac cone is formed with massless and massive dispersions along perpendicular directions. Here we show direct evidence of the highly anisotropic transport of polaritons in a honeycomb lattice of coupled micropillars implementing a semi-Dirac cone. If we optically induce a vacancy-like defect in the lattice, we observe an anisotropically localized polariton distribution in a single sublattice, a consequence of the semi-Dirac dispersion. Our work opens up new horizons for the study of transport and localization in lattices with chiral symmetry and exotic Dirac dispersions.

\end{abstract}
\date{\today}%
\maketitle

Graphene presents extraordinary transport properties arising from the particular conical structure of its Dirac cones. At the Dirac energy, electrons behave as chiral relativistic particles with no mass~\cite{Novoselov2005}. Remarkable effects arise from this unusual electronic band structure, such as, conical diffraction~\cite{Peleg2007}, integer quantum Hall effect at room temperature~\cite{Novoselov2007}, antilocalization~\cite{Wu2007} and Klein tunneling~\cite{Stander2009, Young2009, Allain2011, Ozawa2017}. While the Dirac cones in graphene are cylindrically symmetric, anisotropic Dirac cones in strained two-dimensional materials have driven much attention due to the possibility of modifying the Fermi surface and implementing directional transport properties. For instance, by tuning the nearest-neighbor and next-nearest-neighbor hoppings between atoms along a given direction in tight-binding models, it has been shown that tilted Dirac cones with asymmetric Dirac velocities in the $x$ and $y$ directions can be engineered~\cite{Park2008, Goerbig2008, Wunsch2008, Dietl2008, Pereira2009, Montambaux2009, Montambaux2009b, DeGail2012, Ibanez-Azpiroz2013, Feilhauer2015}. They have been predicted to show exotic tunneling properties~\cite{Volovik2016, Nguyen2017} and high temperature superconducting gaps~\cite{Li2017b}.


A peculiar case of Dirac cone manipulation takes place in a honeycomb lattice when two topologically nonequivalent Dirac cones merge in the presence of uniaxial strain~\cite{Hasegawa2006, Zhu2007, Wunsch2008, Montambaux2009}. In this case, quasiparticles at the Dirac point behave as massless particles in one spatial direction and as massive ones in the perpendicular direction, in a so-called semi-Dirac cone. The asymmetry of such exotic Dirac cones anticipates highly anisotropic transport and localization properties as studied in a number of theoretical works~\cite{Hasegawa2006, Zhu2007, Wunsch2008, Montambaux2009, Dutreix2013,Lim2012, Adroguer2016}. However, these properties have been hardly explored experimentally due to the difficulty in synthesizing two-dimensional materials with the required asymmetric hoppings and low disorder. For instance, semi-Dirac cones have been observed in black phosphorous~\cite{Kim2015a}, but no transport studies are available. Artificial systems, such as ultracold atoms~\cite{Tarruell2012}, lattices of photonic resonators~\cite{Bellec2013, Bellec2014} and waveguide arrays~\cite{Rechtsman2013} have shown the possibility of engineering semi-Dirac cones with an exquisite control, and demonstrated the effect of the merging of the Dirac cones on the presence of edge states~\cite{Rechtsman2013, Bellec2014}. However, transport and localization properties have not been studied in these artificial systems because of the need to access simultaneously spectral information and particle dynamics.

\begin{figure*}[t]
\begin{center}
  \includegraphics[width=\textwidth]{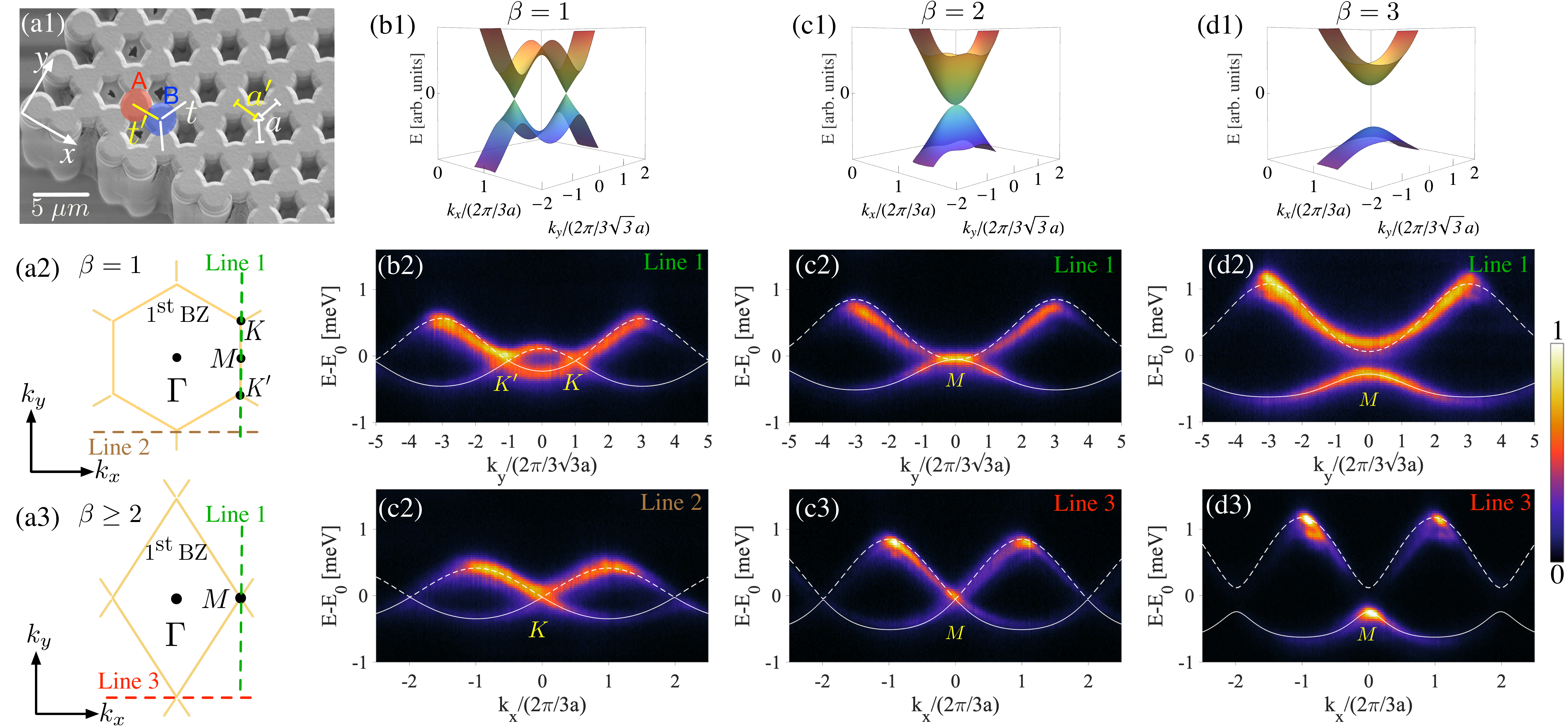}
  	\caption{\label{fig1}  Dirac cone merging and semi-Dirac dispersion. (a1) Electron microscopy image of a polariton honeycomb lattice. Red and blue circles demarcate A and B sublattices. Yellow and white lines denote hopping among horizontal and diagonal nearest neighbors ($t'$ and $t$, respectively). (a2)-(a3) Sketch of the Brillouin zones in momentum space for $\beta=1$ and $\beta \geq 2$.
  	Middle and bottom rows of columns (b)-(d): measured polariton photoluminescence intensity in momentum space for different values of $\beta$. Each image is normalized to its maximum intensity. (b2), (c2), (d2) display the emission along $k_y$ for $k_x=2\pi/3a$ [Line 1 in (a2) and (a3)], while (b3), (c3), (d3) exhibit the measurements along $k_x$ for $k_y=-4\pi/3\sqrt{3}a$ (Line 2), and for $k_y=-6\pi/3\sqrt{3}a$ (Line 3). The white continuous and dashed lines are fits to the lower and upper tight-binding bands [Eq.~(\ref{Bands})]. \mbox{$E_0=1589.2$ meV} and $a=2.4\,\mu$m.}
   \end{center}
 \end{figure*} 
 
In this letter, we experimentally report the highly anisotropic transport and localization properties of polaritons in lattices of semiconductor micropillars~\cite{Jacqmin2014,Klembt2018,Milicevic2019} showing a semi-Dirac dispersion. 
We reveal the anisotropic transport of polaritons along perpendicular spatial directions with massive and massless dispersions, characteristic of the semi-Dirac cone. Taking advantage of the driven-dissipative nature of polaritons we induce effective lattice vacancies, which result in localized polariton distributions with an anisotropic decay and confined in a single honeycomb sublattice. Our observations reveal clear evidence of the long-sought anisotropic transport in unconventional Dirac cones, and provide a new route to implement a localized response in lattices with chiral symmetry.

To engineer the semi-Dirac cone Hamiltonian, we employ lattices of semiconductor micropillars. The lattices are fabricated from a planar semiconductor microcavity made of 28 (top) and 40 (bottom) pairs of $\lambda$/4 alternating layers of Ga$_{0.05}$Al$_{0.95}$As and Ga$_{0.80}$Al$_{0.20}$As ($\lambda= 783$~nm), a $\lambda$/2 cavity spacer of Ga$_{0.05}$Al$_{0.95}$As, and twelve GaAs quantum wells embedded at the three central maxima of the electromagnetic field. At 10 K, the temperature of our experiments, the microcavity is in the strong coupling regime between quantum well excitons and confined photons, giving rise to polaritons characterized by a Rabi splitting of 15 meV. The microcavity is then etched down to the substrate into honeycomb lattices of coupled micropillars of $2.6\;\mu$m diameter. 
By varying the center-to-center distance between micropillars, the amplitude of the polariton hopping between neighboring micropillars can be engineered~\cite{Milicevic2019} to simulate the homogeneous strain that has been predicted to result in semi-Dirac dispersions~\cite{Hasegawa2006, Zhu2007, Wunsch2008, Montambaux2009}.
All experiments are done at a photon-exciton detuning of $-15.2$~meV, thus leading to polaritons states with a dominant photonic fraction, which present the longest polariton lifetimes in our samples.

Figure~\ref{fig1}(a1) shows a scanning electron microscope image of a lattice with isotropic hoppings, corresponding to a center-to-center distance of $a=a'=2.4~\mu$m for the three nearest-neighbors links of each micropillar. To measure the polariton dispersion and study the transport properties, photoluminescence experiments are done under excitation at the center of the lattice in a spot of $8\; \mu$m with a continuous wave laser at $745$~nm. A detailed description of the experimental set-up can be found in Ref.~\cite{Supplemental}. Figure~\ref{fig1}(b2)-(b3) shows the emission from the lowest energy bands (\textit{s}-bands) in momentum space. Along the $k_y$ direction [line 1 in Fig.~\ref{fig1}(a2)] two Dirac crossings are observed in Fig.~\ref{fig1}(b2), corresponding to the \textit{K} and \textit{K'} points characteristic of the unperturbed honeycomb lattice. The Dirac velocities (slopes of the Dirac dispersion) are in this case isotropic around each Dirac cone, as evidenced when comparing the dispersions close to $E_0$ in Fig.~\ref{fig1}(b2) for \textit{K} along $k_y$ [Line 1 in Fig.~\ref{fig1}(a2)] and in Fig.~\ref{fig1}(b3) along $k_x$ [Line 2 in Fig.~\ref{fig1}(a2)].

The polariton dispersion is well reproduced by a tight-binding model
 whose eigenvalues are~\cite{CastroNeto2009, Jacqmin2014}:
 
  \begin{equation}
\label{Bands}
E_{\pm}({\bf k})=E_{0}\pm t\sqrt{(\beta^2 +2)+f({\bf k})} -\bar{t}f({\bf k})\,,
\end{equation} 

\noindent with, $f({\bf k})=2\cos\left(\sqrt{3}k_ya\right)+4\beta\cos\left(\frac{3}{2}k_xa\right)\cos\left(\frac{\sqrt{3}}{2}k_ya\right)$, and $E_{0}$ the Dirac-point energy. $t$ and  $\bar{t}$ refer to nearest- (NN) and next-nearest-neighbor (NNN) hoppings, respectively, while $\beta\equiv t'/t$ represents the ratio of the horizontal polariton hopping to the diagonal one [Fig.~\ref{fig1}(a1)]. Hence, $\beta$ quantifies the engineered compression strength, which is equal to 1 in the present case (isotropic hopping). A fit of Eq.~(\ref{Bands}) to the collected photoluminescence [white lines in Fig.~\ref{fig1}(b2)-(b3)] results in the hopping parameters $t=0.18$~meV and $\bar t=-0.014$~meV. Note that in the micropillar system, the NNN hopping in Eq.~(\ref{Bands}) is a phenomenological term that reproduces the observed asymmetry of \textit{s}-bands. Its origin is the coupling of \textit{s} and \textit{p}-modes, as described in Ref.~\cite{Mangussi2020}.

\begin{figure}[t!]
\begin{center}
  \includegraphics[width=0.42\textwidth]{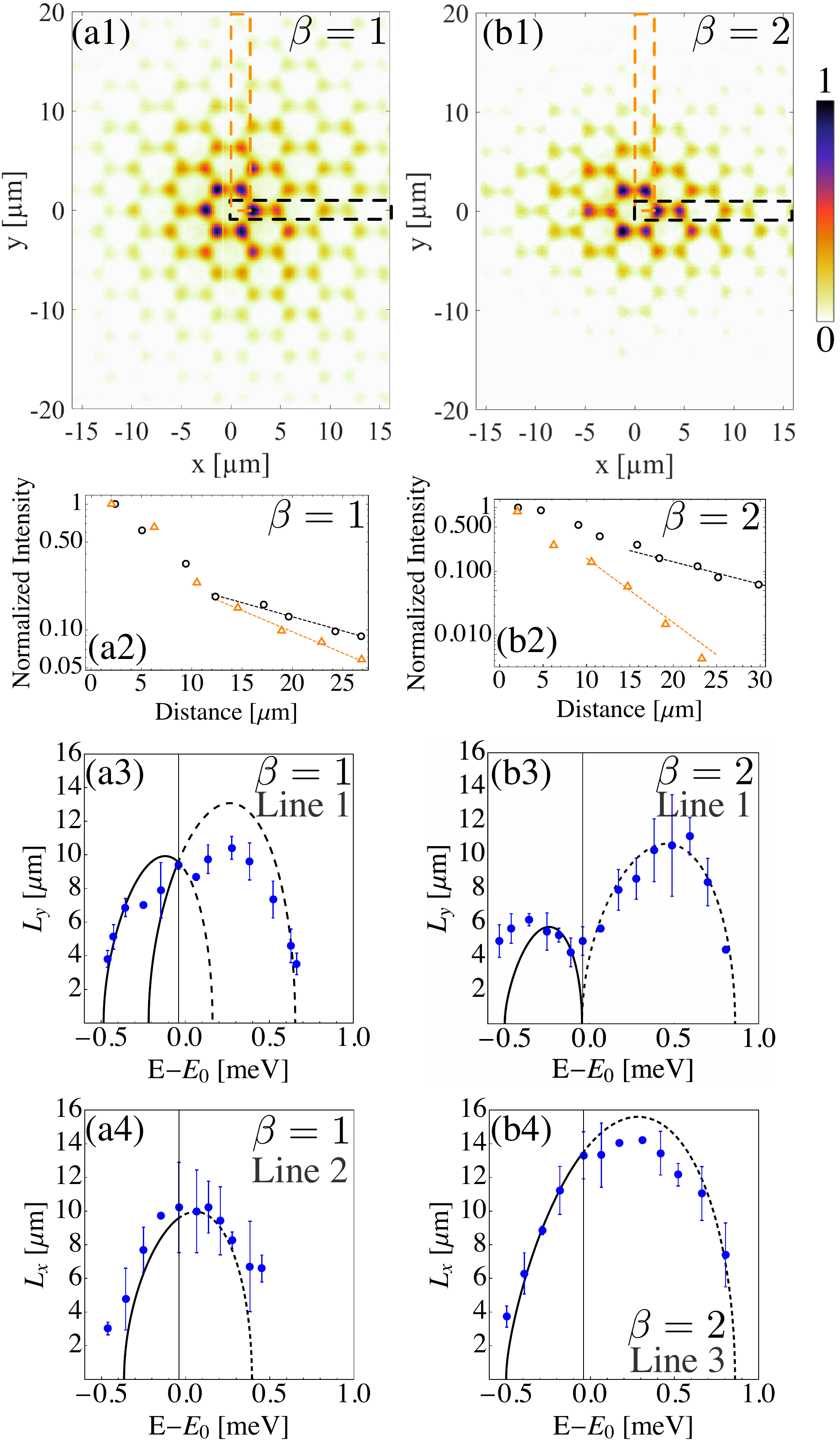}
  \caption{\label{fig2} Transport at the Semi-Dirac point. (a1) and (b1) show the photoluminescence intensity in real space at the energy of the Dirac point ($E_0=1589.2$ meV) for $\beta=1$ and $\beta=2$, respectively. Each image is normalized to its maximum intensity. (a2)-(b2) show the measured intensity along $x$ (circles) and $y$ (triangles) directions for $\beta=1$ (a2) and $\beta=2$ (b2), extracted from the dashed boxes in (a1) and (b1). Lines are exponential decay fits. (a3) and (a4) present the measured propagation lengths (dots), at several energies, along $y$ and $x$ directions for $\beta =1$. (b3) and (b4): same for $\beta =2$. Solid and dashed lines display the theoretical propagation lengths [Eq.~(\ref{PL})] corresponding to the solid and dashed bands in Fig.~\ref{fig1}(b2),(b3),(c2),(c3). 
  The vertical line depicts the Dirac-point energy $E_0$.}
   \end{center}
 \end{figure}

When the horizontal hopping $t'$ is increased, the tight-binding bands show that the Dirac cones \textit{K} and \textit{K'} move towards each other, and for a value of $\beta=2$ they merge at a single point [Fig.~\ref{fig1}(c1)]. We experimentally probe this situation in Fig.~\ref{fig1}(c2)-(c3) for a lattice with $a'=2.2\;\mu$m and $a=2.4\;\mu$m. Using the tight-binding model with the previously obtained values of $t$ and $\bar t$, a value of $\beta=2$ reproduces the experimental features. The recorded spectrum along $k_y$ [Line 1 in Fig.~\ref{fig1}(a3)] shows not only that the two Dirac cones have merged but, more importantly, the dispersions of both the upper and lower bands are now parabolic in this direction, while they remain linear along the $k_x$ direction. This situation is known as a semi-Dirac cone, which combines massless and massive dispersions along perpendicular directions. They have been observed in ARPES measurements of strained black phosphorus~\cite{Kim2015a} and indirectly in various artificial lattices~\cite{Tarruell2012,Bellec2013, Bellec2014, Rechtsman2013}. If $\beta$ is further increased, the Dirac cone merging evolves into a band gap [see Fig.~\ref{fig1}(d1)]. We implement experimentally this situation by reducing further the center-to-center distance $a'$ to $1.7\;\mu$m as shown in Fig.~\ref{fig1}(d2)-(d3), corresponding to $\beta=3$. 

The anisotropic dispersion of the semi-Dirac cone for $\beta=2$ is expected to have strong consequences in the transport properties of polaritons. To study this effect, we probe the polariton distribution in real space at different emission energies. Figure~\ref{fig2}(a1) and (b1) show the real-space intensity at the energy $E_0$ of the Dirac point for $\beta=1$ and $\beta=2$, respectively. For $\beta=1$ [panel (a1)], we observe that polaritons travel away from the excitation spot isotropically; on the contrary for $\beta=2$ [panel (b1)], the propagation is significantly anisotropic, being more pronounced in the \textit{x} direction than in the \textit{y} direction. To quantify this anisotropy, we measure the propagation length on both $x$ and $y$ directions at $E_0$ in both lattices. The propagation length is extracted by fitting an exponential decay to the tails of the emitted intensity, i.e. $|\psi(r)|^2\propto e^{-r/L_r}$ along $x$ and $y$ directions [enclosed region in Fig.~\ref{fig2}(a1) and (b1)]. Experimental points and fits are shown in Fig.~\ref{fig2}(a2) and (b2). For $\beta=1$, the propagation lengths are $L_{x}=10.21\,\pm\,2.69\;\mu$m and $L_{y}=9.38\,\pm\,0.23\;\mu$m. These values confirm the isotropic transport of polaritons near the Dirac-point energy, which was previously measured in the form of conical diffraction~\cite{Peleg2007}. For $\beta=2$, at the same energy, we obtain $L_{x}=13.31\,\pm\,1.40\;\mu$m and $L_{y}=4.89\,\pm\,0.84\;\mu$m, evidencing the high group velocity in the direction of the massless dispersion, and the reduced group velocity along the $y$ direction associated to the touching parabolic bands.

Figure~\ref{fig2}(a3)-(b4) shows in filled dots the measured propagation lengths $L_x$ and $L_y$ as a function of the energy across the Dirac point. 
This measurement can be directly compared to the propagation length expected from the group velocities, $v_{g,x(y)}=\partial E / \partial k_{x(y)}$, in the following way:
\begin{equation}
\label{PL}
L_{x(y)}\approx v_{g,x(y)} \tau .
\end{equation}
$v_{g,x(y)}$ is calculated from the dispersion curves in Fig.~\ref{fig1} along the vertical ($k_x=2\pi / 3a$) and horizontal ($k_y=-4\pi/3\sqrt{3}a$ for $\beta=1$; $k_y=-6\pi/3\sqrt{3}a$ for $\beta=2$) directions, and $\tau$ is the polariton lifetime. The lines in Fig.~\ref{fig2}(a3)-(b4) show the propagation lengths calculated from the group velocities predicted by the tight-binding model in each spatial direction below (continuous line) and above (dashed line) $E_0$. Here we assume a polariton lifetime of $\tau=14$~ps and $\tau=12$~ps, for $\beta=1$ and $\beta=2$ lattices, respectively, which is used as a fitting parameter to the experimental points.

\begin{figure}[t]
\begin{center}
  \includegraphics[width=0.40\textwidth]{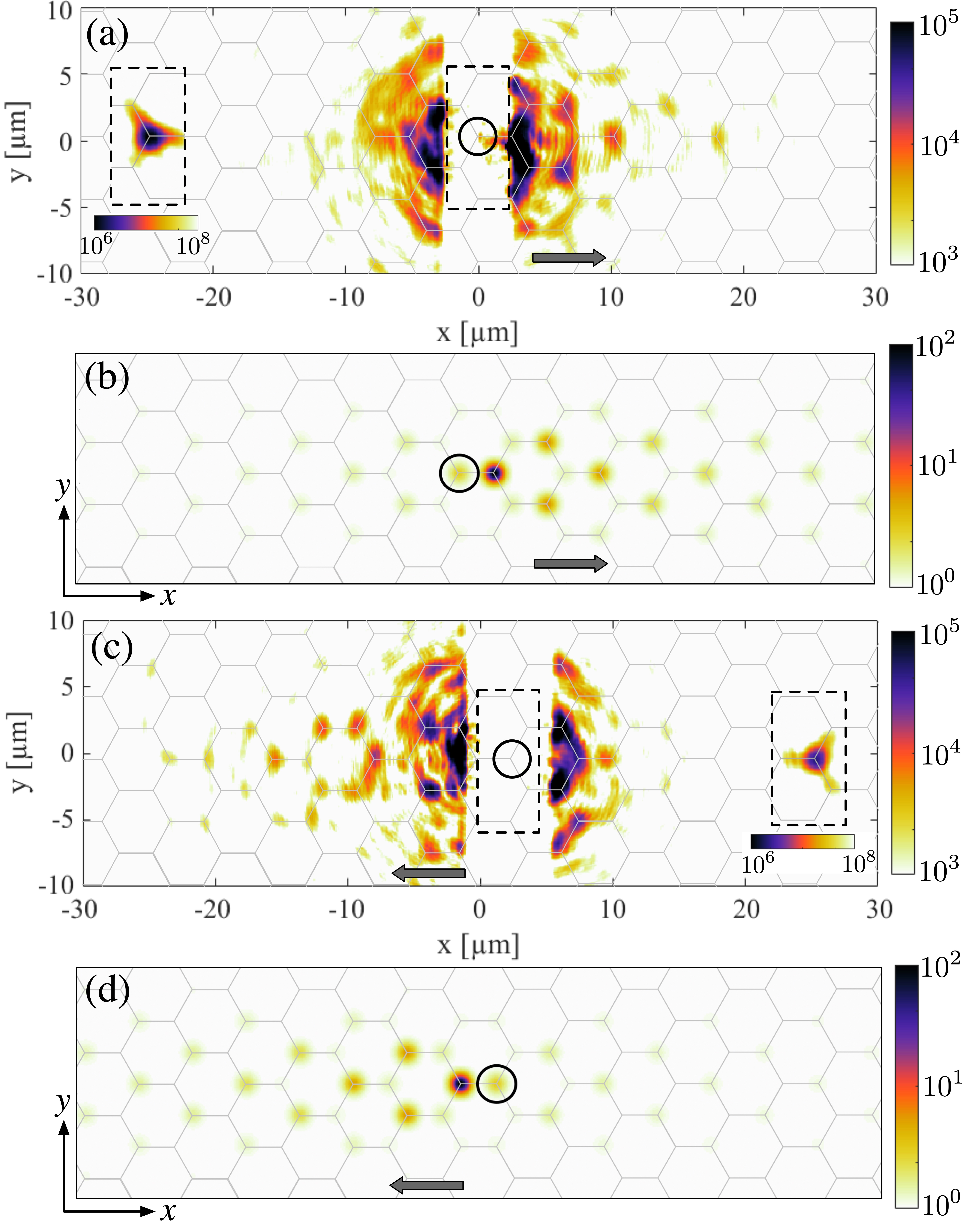}
  \caption{\label{fig3}All-optical analog of a vacancy localization in semi-Dirac graphene. (a) and (b) show the measured photoluminescence intensity in the real space at the energy of the Dirac point when a single pillar is pumped (demarcated by a circle) in an A pillar (a) and in a B pillar (c). Insets show the reflected pump spot when the beam block is removed from the central region. (b) and (d) show the polariton distribution calculated from Eq.~(\ref{eq:Schrod}) when a single A and B pillar, respectively, is pumped at $E_0$ energy. Hexagons depict the underlying lattice.} 
   \end{center}
 \end{figure}
 
The calculated propagation distances match well the experimental data and reproduce the increase of the propagation length along the $x$ direction when going from $\beta=1$ to $\beta=2$, due to the higher hopping in that direction [see Fig.~\ref{fig2}(a4)-(b4)]. Along the $y$ direction, the expected propagation length for $\beta=2$ goes down to zero at the Dirac-point energy $E_0$, a consequence of the massive dispersion along that direction [see Fig.~\ref{fig2}(a3)]. Similarly, the calculated propagation length is also zero at the top and bottom of the bands. Experimentally, the measured propagation length at those points is about $4\;\mu$m. This value is, in part, determined by the linewidth of 0.060~meV associated to the finite polariton lifetime: when selecting a given energy, we are in fact detecting the emission from a small range of energies around the desired one, corresponding to states with a nonzero group velocity. Moreover, diffusion of photoexcited excitons away from the excitation spot might also contribute to the residual measured propagation distance. 

Further insights on the transport properties at the semi-Dirac cone energy  $E_0$ can be accessed when implementing a resonant-laser excitation scheme. Figure~\ref{fig3}(a) shows the measured intensity when a resonant laser at $E_{0}$ is focused on a single micropillar of the A sublattice (marked with a circle) in a lattice with $\beta=2$. To measure the propagation away from the excitation spot, a mask was placed at the center of the image (white area) with the aim of blocking the excitation beam reflecting onto the CCD (the inset shows an image of the reflected pump beam in the absence of the mask). The image shows some stray laser light close to the excitation spot and a decay of the luminescence on the B sublattice towards the right of the excitation spot. If the excitation is centered on a pillar of the B sublattice, the decay direction and the sublattice asymmetry are reversed, as shown in Fig.~\ref{fig3}(c). Further data analysis can be found in Supplemental Material.

This behavior is well reproduced using a driven-dissipative model of the polariton dynamics in resonant excitation~\cite{Carusotto2013}:
\begin{equation}
i\hbar\frac{\partial \psi_{n}}{\partial t}=\sum_{m \neq n}t_{n, m}\psi_{m}-i \frac{\hbar }{\tau}\psi_{n}\, +F\delta _{n, n_p}e^{i \omega_pt}.
\label{eq:Schrod}
\end{equation}

\noindent $\psi_{n}$ represents the polariton amplitude at site $n$, $t_{n, m}$ is the nearest-neighbor hopping, and $F$ is the strength of the pump at frequency $\omega_p$. Figure~\ref{fig3}(b) depicts the steady-state solution in the conditions of Fig.~\ref{fig3}(a): $\tau=12$~ps, $t=0.18$~meV, $\beta=2$. It shows that the population in the pumped micropillar, marked by a circle, is almost zero, and the distribution extends mainly to the right of the excited micropillar, on the B sublattice. When moving the excitation spot to a B site [Fig.~\ref{fig3}(d)], the distribution reverses its decay direction, as observed in the experiment [panel (c)]. Note that along the \textit{y} direction, corresponding to the massive dispersion of the semi-Dirac point, the polariton distribution is localized within a single hexagon.

The observed polariton distributions resemble the predicted wavefunction of electrons bound to a single bulk vacancy in compressed graphene~\cite{Dutreix2013}. It has been shown that a single bulk vacancy in graphene creates a defect state at the Dirac-point energy $E_{0}$, with a decay in amplitude following a $1/r$ law~\cite{Pereira2006, Wehling2007, Dutreix2013}. The chiral symmetry of the lattice imposes that its wavefunction resides in one sublattice only: the sublattice opposite to that of the vacancy. Experiments shown in Supplemental Material for lattices with $\beta=1$ reproduce this situation. In the case of a semi-Dirac cone, the wavefunction of the vacancy state acquires an anisotropic distribution: if the vacancy is in the A sublattice, the state is localized to the right of the vacancy; if the vacancy is in the B sublattice, it is localized to the left~\cite{Dutreix2013}. In both cases the decay of the amplitude follows $1/\sqrt{|x|}$. This vacancy states are expected to play an important role in the transport properties of graphene-like materials in which localization by weak disorder is strongly decreased due to the Klein tunneling effect.


The similarity between the measured polaritonic distribution and bound electron wavefunctions can be interpreted phenomenologically as follows. Under resonant excitation ($\hbar \omega_p=E_{0}$), the population of the driven micropillar interferes destructively with the laser, resulting in an almost zero population in the pumped micropillar, analogous to the effect of a vacancy. This phenomenon was recently reported in the case of two coupled micropillars~\cite{Rodriguez2016}, and it is expected to happen in any lattice of micropillars with chiral symmetry.

In summary, we have probed the simultaneous massive and massless behavior of polaritons in a semi-Dirac honeycomb lattice. Additionally, we have generated an all-optical analog of a vacancy and reported the associated anisotropic distribution in the bulk of the lattice. 
The use of polariton nonlinearities, absent in the experiments presented here, are a promising perspective for the study of nonlinear modes at Dirac and semi-Dirac points~\cite{Chen2011}, hardly studied so far due to the rarity of systems with engineered Dirac cones and nonlinearities.

\noindent \textit{Acknowledgements} - We thank J.-N. Fuchs, F. Mangussi and G. Usaj for fruitful discussions. This work was supported by the H2020-FETFLAG project PhoQus (820392), the QUANTERA project Interpol (ANR-QUAN-0003-05), the French National Research Agency project Quantum Fluids of Light (ANR-16-CE30-0021), the Paris Ile-de-France Region in the framework of DIM SIRTEQ, the Marie Sklodowska-Curie individual fellowship ToPol, the Labex NanoSaclay (ANR-10-LABX-0035), the French government through the Programme Investissement d’Avenir (I-SITE ULNE / ANR-16-IDEX-0004 ULNE) managed by the Agence Nationale de la Recherche, the French RENATECH network, the Labex CEMPI (ANR-11-LABX-0007), the CPER Photonics for Society P4S and the M\'etropole Europ\'eenne de Lille (MEL) via the project TFlight. T.O. is supported by JSPS KAKENHI Grant Number JP18H05857, JST PRESTO Grant Number JPMJPR19L2, JST CREST Grant Number JPMJCR19T1, and the Interdisciplinary Theoretical and Mathematical Sciences Program (iTHEMS) at RIKEN.


\bibliography{references}{}
\end{document}


\title{Supplemental Material: Semi-Dirac transport and anisotropic localization in polariton honeycomb lattices}%

\author{B.~Real}
\affiliation{Univ. Lille, CNRS, UMR 8523 -- PhLAM -- Physique des Lasers Atomes et Mol\'ecules, F-59000 Lille, France}

\author{O.~Jamadi}
\affiliation{Univ. Lille, CNRS, UMR 8523 -- PhLAM -- Physique des Lasers Atomes et Mol\'ecules, F-59000 Lille, France}

\author{M.~Mili\'cevi\'c}
\affiliation{Centre de Nanosciences et de Nanotechnologies~(C2N),CNRS -~Universit\'e Paris-Sud/Paris-Saclay, Palaiseau, France}

\author{N.~Pernet}
\affiliation{Centre de Nanosciences et de Nanotechnologies~(C2N),CNRS -~Universit\'e Paris-Sud/Paris-Saclay, Palaiseau, France}

\author{P.~St-Jean}
\affiliation{Centre de Nanosciences et de Nanotechnologies~(C2N),CNRS -~Universit\'e Paris-Sud/Paris-Saclay, Palaiseau, France}

\author{T.~Ozawa}
\affiliation{Advanced Institute for Materials Research, Tohoku University, Sendai 980-8577, Japan}
\affiliation{Interdisciplinary Theoretical and Mathematical Sciences Program (iTHEMS), RIKEN, Wako, Saitama 351-0198, Japan}

\author{G.~Montambaux}
\affiliation{Laboratoire de Physique des Solides, CNRS, Universit\'e Paris-Sud, Universit\'e Paris-Saclay, 91405 Orsay Cedex, France}

\author{I.~Sagnes}
\affiliation{Centre de Nanosciences et de Nanotechnologies~(C2N),CNRS -~Universit\'e Paris-Sud/Paris-Saclay, Palaiseau, France}

\author{A.~Lema\^itre}
\affiliation{Centre de Nanosciences et de Nanotechnologies~(C2N),CNRS -~Universit\'e Paris-Sud/Paris-Saclay, Palaiseau, France}

\author{L.~Le Gratiet}
\affiliation{Centre de Nanosciences et de Nanotechnologies~(C2N),CNRS -~Universit\'e Paris-Sud/Paris-Saclay, Palaiseau, France}

\author{A.~Harouri}
\affiliation{Centre de Nanosciences et de Nanotechnologies~(C2N),CNRS -~Universit\'e Paris-Sud/Paris-Saclay, Palaiseau, France}

\author{S.~Ravets}
\affiliation{Centre de Nanosciences et de Nanotechnologies~(C2N),CNRS -~Universit\'e Paris-Sud/Paris-Saclay, Palaiseau, France}

\author{J.~Bloch}
\affiliation{Centre de Nanosciences et de Nanotechnologies~(C2N),CNRS -~Universit\'e Paris-Sud/Paris-Saclay, Palaiseau, France}

\author{A.~Amo}
\affiliation{Univ. Lille, CNRS, UMR 8523 -- PhLAM -- Physique des Lasers Atomes et Mol\'ecules, F-59000 Lille, France}

%
%
\date{\today}%
\maketitle

\section{Setup and Experimental conditions}
\begin{figure}[b]
\begin{center}
  \includegraphics[width=0.8\textwidth]{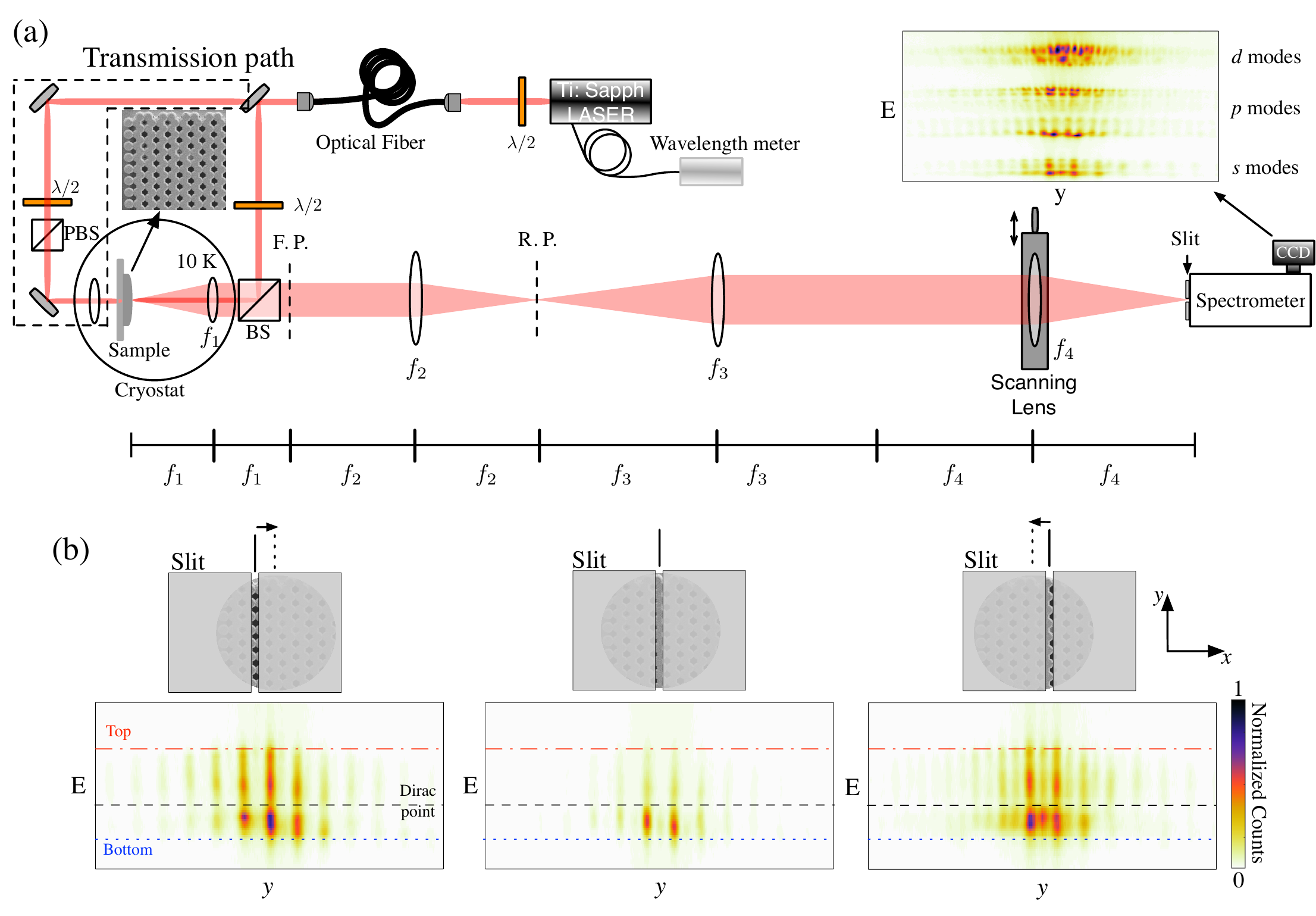}
  \caption{\label{figSuppl1}  (a) Scheme of the experimental setup to measure the polariton emission at a given energy. The region in the dashed rectangle shows the beam path for transmission experiments. (b) The upper row depicts selected regions of the image of the lattice by a slit at three different horizontal (\textit{x}) positions. The lower row presents the population of polaritons as a function of the energy (\textit{E}) and the real space along vertical (\textit{y}) direction for the respective horizontal positions. Dotted, dashed and dot-dashed lines respectively point out the bottom, Dirac-point and top energy of the s bands. } 
   \end{center}
 \end{figure}

The experiments reported in the main text are done using the experimental setup schematized in Fig.~\ref{figSuppl1}(a). The experiments are carried out at low temperature ($10$ K) by placing the sample (lattices of semiconductor micropillars) in a closed-cycle cryostat. We excite non resonantly the lattices by using a linearly polarized beam coming from a continuous-wave monomode Ti:Sapphire laser at $745$ nm. After traveling through a polarization-maintaining monomode optical fiber, the beam shows a clear gaussian profile. From there, the beam is driven to the cryostat where it is tightly focused on the lattices by an aspherical lens of $8$-mm focal length (NA$=0.45$) [$f_1$ in Fig.~\ref{figSuppl1}(a)]. 
The absorption of the laser by the semiconductor creates a hot cloud of electron-hole pairs that incoherently relax and populate all the polaritonic bands of the structure. Since polaritons have a finite lifetime (12-14~ps) they leak out the cavity in the form of photons that encode the energy and momentum of the polaritons inside the structure. This photoluminescence is collected by the same $8$-mm lens used for the excitation and an image system generates a 120-times magnified image of the lattice at the entrance port of a spectrometer. The entrance slit of the spectrometer, and a scanning lens placed on a motorized translation stage [see Fig.~\ref{figSuppl1}(a)] allow selecting the real space image of a vertical slice of the emission. The slit has a width of $\sim 32.5\,\mu$m which, taking into account the magnification of the set-up, corresponds to $0.3\,\mu$m on the sample [see Fig.~\ref{figSuppl1}(b) upper row]. 
This $1$D $y$-slice of the image is dispersed by the spectrometer and imaged on a CCD. The image of the CCD, shown in Fig.~\ref{figSuppl1}(b) lower row, represents $y$ position in one direction and energy on the other. Thus, the energy-resolved emission from the polariton bands can be recorded along \textit{y} direction for a specific \textit{x} position. Using the motorized-translation stage, the position of the scanning lens can be shifted in the $x$ direction such that the emission from different $x$ positions on the lattice can be recorded. 
From all the recorded spectra, real space images of the emission at any given energy can be reconstructed. The energy precision is given by the pixel resolution of the CCD, being $33.1\,\mu$eV.

Additionally, we also perform an angle-resolved scan of the momentum space by removing $f_3$ lens. By doing so, the imaging system formed by the $f_2$ lens and the scanning lens images the Fourier plane of $f_1$ on the entrance port of the CCD. Each point in this plane corresponds to an angle of emission of the sample $\theta$, which is directly related to the in-plane momentum of polaritons $k_{|}$ via $k_{|}=k_0 \cos(\theta)$, where $k_0$ is the total momentum of the out coming photon. This configuration of the imaging system allows us to observe the band structure for given cuts of the Fourier plane as shown in Fig. 1 of the main text. Therefore, we can both image the momentum and real space of the emission. 

To perform resonant experiments, we set the wavelength of the laser to the (semi-)Dirac-point energy with the help of a wavelength meter. 
The energy of the (semi-)Dirac point is obtained from the spectra recorded in the non-resonant experiment described above. Once the wavelength is set, we focalise the laser spot on one pillar [A or B Fig. 1(a) main text], with FWHM of $\approx 2\,\mu$m. To prevent the intense reflected laser beam from reaching the CCD and, thus, be able to record the emission from the injected polaritons [Fig.~3(a) and (c) main text], we place a spatial filter in the first real plane (R.P.) shown in Fig.~\ref{figSuppl1}(a). The effective size of the spatial filter corresponds to the white central area in Fig.~3(a) and (c) of the main text.
 
 \section{Anisotropic emission from vacancy-like polariton distribution}
 
 \begin{figure}[b]
\begin{center}
  \includegraphics[width=0.7\textwidth]{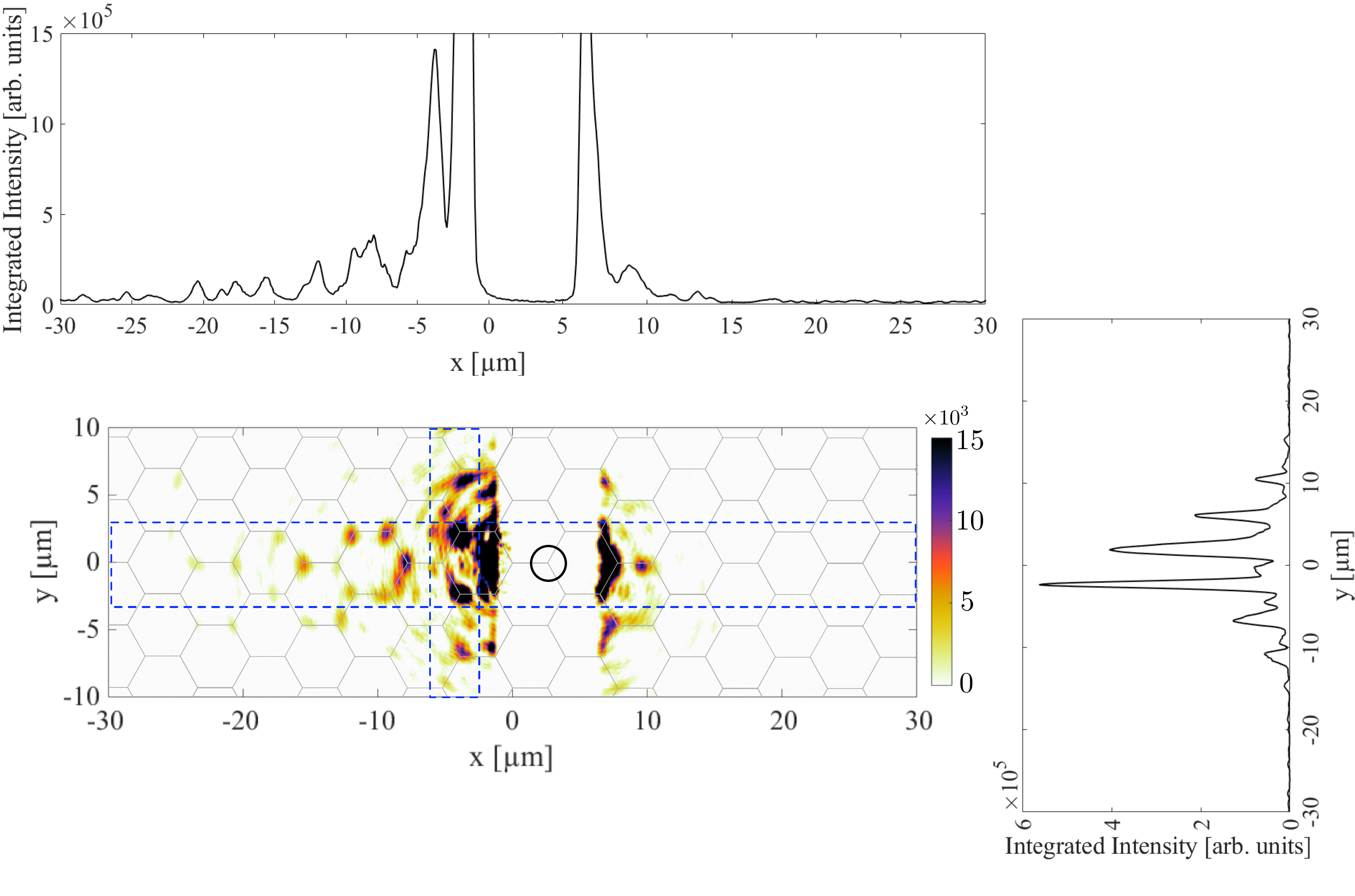}
  \caption{\label{figSuppl3} The central panel shows Fig.~3(c) of the main text in linear color scale. The circle points out the pumped pillar. Top panel shows a vertical integrated plot along the \textit{x} direction, while the right panel shows a horizontal integrated plot along the \textit{y} direction. Enclosed regions depict the integrated areas.} 
   \end{center}
 \end{figure}

 Figure 3 of the main text presents the emission when exciting a lattice with $\beta=2$ at the semi-Dirac cone energy ($E_0$) on a single pillar. As discussed in the main text, the interference between the excitation field the polariton emission results in a near-zero field amplitude at the pillar on which the laser is focused, resulting in the analog of a bulk vacancy. Reference~\cite{Dutreix2013} predicted that vacancy states at $E_0$ should be strongly asymmetric, as clearly seen in Fig.~3 of the main text, which is displayed in logarithmic scale. 
 To provide further support to these observations, the central panel of Fig.~\ref{figSuppl3} shows in linear color scale the data displayed in Fig.~3(c) of the main text, under steady-state excitation of a B site. The directionality of the decay of the emission is further evidenced in the horizontal and vertical profiles shown in the upper and right panels. They show the emitted intensity integrated along the vertical (upper panel) and horizontal (right panel) direction within the dashed rectangles depicted in the central panel. 
 Along the \textit{x} direction, the integrated profile clearly presents a much longer decay towards negative values. 
 Conversely, the right panel exhibits a symmetric and fast decay in the \textit{y} direction. 
 Along this direction we expect the polariton distribution to cover only one hexagon (simulated images in Fig.~3 of the main text). However, scattered light of the reflected pumped beam is also present at longer distances.          
 
 \section{Vacancy states in regular honeycomb lattices ($\beta=1$)}

\begin{figure}[h!]
\begin{center}
  \includegraphics[width=0.7\textwidth]{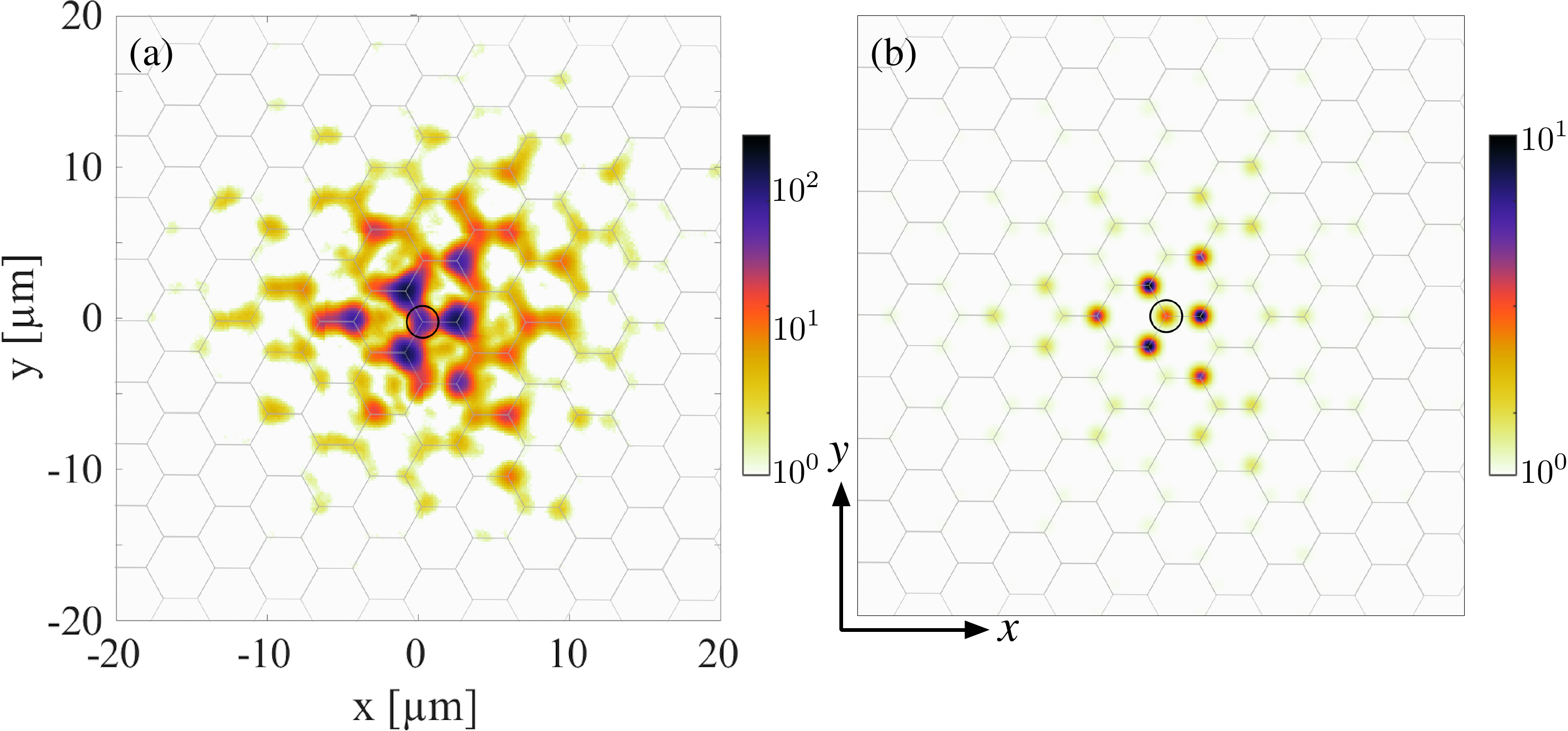}
  \caption{\label{figSuppl2} All-optical analog of a vacancy state in a honeycomb lattice with $\beta=1$. (a) Measured photoluminescence intensity in the real space at the energy of the Dirac point ($E_0$) when a single A-pillar is pumped (marked by a circle). (b) Steady-state solution of Eq. (3) of the main text when a single A-pillar is pumped at $E=E_0$ and considering $t=0.23$ meV, $\beta=1$, and $\tau\approx 10$ ps. } 
   \end{center}
 \end{figure}

We have also explored the formation of vacancy-like states in the bulk of regular honeycomb lattices with $\beta=1$. For this purpose, we fabricate a planar semiconductor microcavity made of $28$ (top) and $32$ (bottom) pairs of $\lambda/4$ alternating layers of Ga$_{0.10}$Al$_{0.90}$As and Ga$_{0.05}$Al$_{0.95}$As ($\lambda =880$ nm), a $\lambda$ spacer of GaAs, and a single $20$-nm-width In$_{0.09}$Ga$_{0.91}$As quantum well at the center of the cavity. This microcavity is etched down to the substrate into honeycomb lattices of coupled micropillars of $2.75\, \mu$m, with a center-to-center distance of $2.4\,\mu$m. At cryogenic temperature the lattices possess a Rabi splitting of $3.5$ meV. The In$_{0.09}$Ga$_{0.91}$As quantum well used in this microcavity sample allows for experiments in transmission geometry, in which the resonant excitation beam impinges on the sample on one side and observation is done from the other side [the excitation beam follows the path of the enclosed region in Fig.~\ref{figSuppl1}(a)]. Fig.~\ref{figSuppl2}(a) presents the measured intensity in logarithmic scale when a laser at the Dirac-point energy ($E_0=1402.54$ meV) is focused on a single micropillar of the A sublattice (marked with a circle). The intensity distribution decays exhibiting a triangular shape with a clear predominance of the population on the B sublattice, opposite to that of the excitation beam. This polariton distribution resembles the wavefunction of electron for a single bulk vacancy in graphene \cite{Pereira2006,Wehling2007}. In contrast to the situation for a compressed lattice ($\beta=2$), the polariton steady state is here spread over \textit{x} and \textit{y} directions. Note that Figure~3 of the main text is also presented in logarithmic scale. Fig.~\ref{figSuppl2}(b) shows the steady-state solution of Eq.~(3) of the main text with the conditions: $E=E_0$, $t=0.23$ meV, $\beta=1$, and $\tau\approx 10$ ps.  

\newpage
\bibliography{referencesSupplemental}{}